\documentclass[aps,prl,twocolumn,groupedaddress,showpacs]{revtex4}
\usepackage{graphicx}
\usepackage{dcolumn}
\usepackage{bm}

\begin{document}

\title{Comment on ``Quantum Solution to the Arrow-of-Time Dilemma"}

\author{Hrvoje Nikoli\'c}
\affiliation{Theoretical Physics Division, Rudjer Bo\v{s}kovi\'{c}
Institute,
P.O.B. 180, HR-10002 Zagreb, Croatia.}
\email{hrvoje@thphys.irb.hr}

\date{\today}

\begin{abstract}
In this paper we criticize the attempt 
of Maccone [Phys. Rev. Lett. {\bf 103}, 080401 (2009)]  
to solve the problem of the arrow of time.
\end{abstract}
 

\maketitle

Recently, Maccone \cite{macc} has shown that all phenomena which leave
a trail behind (and hence can be studied by physics) are those the entropy
of which increases. From that he concluded that
the second law of thermodynamics is reduced to a mere tautology, 
suggesting that it solves the problem of the arrow of time in physics.
Namely, according to \cite{macc}, one observes
that entropy allways increases with time (or remains constant) 
simply because processes in which entropy decreases with time,
if they exist, cannot be detected.

In this Comment we criticize the conclusion above. To be clear, we agree
with the result of \cite{macc} that an observable physical phenomenon 
must leave a trail and that leaving a trail necessarily increases entropy.
Nevertheless, we argue that this fact does not help much to understand
the origin of the arrow of time. More precisely, we discuss 3 different
problems (where two of them, denoted by 2a) and 2b), are not completely
independent) with the arrow of time that the results of \cite{macc}
do not solve.

{\it 1) The problem of macroscopic entropy-decreasing processes}.---Let us explain
this problem through a simple example from everyday life. The forces of nature
(earthquakes, hurricanes, etc.) can easily destroy a house, but they cannot easily
build one. This is so because a destroyed house has a larger  
entropy than an undestroyed one. (This refers to macroscopic coarse-grained entropy, which 
should be distinguished from thermodynamic entropy which is coarse-grained on a finer level.) If processes in which 
entropy decreases exist, then one should expect the existence of 
processes in which houses are built spontaneously. Indeed, such a process can leave
a trail in another medium (say on the film tape that records it) in which the entropy 
increases, so such an entropy-decreasing process should be observable.
Yet, we do not see such processes in nature and the
results of \cite{macc} cannot explain why.

{\it 2a) The trace-from-the-future problem}.---According to \cite{macc},
an entropy-decreasing transformation cannot leave any trace. However
if, {\it a priori}, both directions of time have completely equal roles in physics,
then this cannot be true. This is because {\em de}creasing is the same
as {\em in}creasing in the opposite time direction. Hence, if 
an entropy-increasing transformation leaves a trace from the past, then
an entropy-decreasing transformation leaves a trace from the future.
Yet, in nature we do not see traces from the future and the
results of \cite{macc} do not explain why.
Of course, the fact that we do not see traces from the future
can be explained from the assumption that
(for some unknown reason) we live in a universe in which entropy 
{\em does} increase with time (this is, indeed, the standard explanation \cite{stand}),
but the idea of \cite{macc} is to avoid that assumption entirely.

{\it 2b) The existential time-arrow problem}.---The problem 2a) can be cheaply
avoided by a ``common sense'' argument that traces from the future cannot exist
because the future does not yet exist. However, such an argument presumes that
the future is not equally real as the past, i.e., that a preferred direction of time
exists. Therefore, we refer to the paradigm according to which the future does not yet exist
as the existential time arrow. In this way, \cite{macc} cannot completely explain
the origin of the arrow of time, but at best can explain the thermodynamic 
time arrow from the assumption of the existential time arrow.
(This should be contrasted with the standard view \cite{stand} according to which 
the existential time arrow is only an illusion emerging from the assumption 
that entropy increases with time.)

To conclude, the results of \cite{macc} do not help much to solve the difficult
problem of the origin of the arrow of time.

This work was supported by the Ministry of Science of the
Republic of Croatia under Contract No.~098-0982930-2864.
\\

\end{document}